\documentclass[aps,pra,twocolumn,superscriptaddress]{revtex4}
\usepackage{hyperref}
\usepackage{braket}
\usepackage{amsmath}
\usepackage{amsfonts}
\usepackage{amssymb}
\usepackage{graphicx}
\usepackage{epsf}
\usepackage[sort&compress]{natbib}
\usepackage{subfigure}
\usepackage{xcolor}
\hypersetup{backref,
	pdfpagemode=FullScreen,
	colorlinks=true, linkcolor=blue, urlcolor=blue, citecolor=blue}
\usepackage{ulem}

\begin{document}
\title{Two-level system as a quantum sensor of absolute power}
\author{T. H\"{o}nigl-Decrinis}
\email[]{teresa.hoenigl-decrinis@npl.co.uk}
\affiliation{Physics Department, Royal Holloway, University of London, Egham, Surrey TW20 0EX, United Kingdom}
\affiliation{National Physical Laboratory, Teddington, TW11 0LW, United Kingdom}

\author{R. Shaikhaidarov}
\affiliation{Physics Department, Royal Holloway, University of London, Egham, Surrey TW20 0EX, United Kingdom}
\affiliation{Moscow Institute of Physics and Technology, 141700 Dolgoprudny, Russia}

\author{S. E. de Graaf}
\affiliation{National Physical Laboratory, Teddington, TW11 0LW, United Kingdom}

\author{V.N. Antonov}
\affiliation{Skolkovo Institute of Science and Technology, Nobel str. 3, Moscow, 143026, Russia}
\affiliation{Physics Department, Royal Holloway, University of London, Egham, Surrey TW20 0EX, United Kingdom}
\affiliation{Moscow Institute of Physics and Technology, 141700 Dolgoprudny, Russia}

\author{O.V. Astafiev}
\affiliation{Skolkovo Institute of Science and Technology, Nobel str. 3, Moscow, 143026, Russia}
\affiliation{Physics Department, Royal Holloway, University of London, Egham, Surrey TW20 0EX, United Kingdom}
\affiliation{National Physical Laboratory, Teddington, TW11 0LW, United Kingdom}
\affiliation{Moscow Institute of Physics and Technology, 141700 Dolgoprudny, Russia}
\date{\today}


\begin{abstract}
A two-level quantum system can absorb or emit not more than one photon at a time. Using this fundamental property, we demonstrate how a superconducting quantum system strongly coupled to a transmission line can be used as a sensor of the photon flux. We propose four methods of sensing the photon flux and analyse them for the absolute calibration of power by measuring spectra of scattered radiation from the two-level system. This type of sensor can be tuned to operate in a wide frequency range, and does not disturb the propagating waves when not in use. Using a two-level system as a power sensor enables a range of applications in quantum technologies, here in particular applied to calibrate the attenuation of transmission lines inside dilution refrigerators.
\end{abstract}
\maketitle

\section{Introduction}

Progress in development of superconducting circuits, in particular applications in quantum optics, quantum computing and quantum information, demand calibration of microwave lines and knowledge of applied powers to the circuits situated on a chip at millikelvin temperatures.
Usually, one resorts to room-temperature characterisation with power meters and spectral analysers based on semiconductor electronics. However, when the setup including several microwave components (wiring, attenuators, circulators, amplifiers, etc.) is cooled down to millikelvin temperatures, their transfer functions are changed. Furthermore, the circuits on chip are usually omitted from room temperature characterisations.

There have been several proposals to tackle this problem. For instance using Planck spectroscopy~\citep{Mariantoni:wh, Goetz:vm}, the shot noise of a known microwave component~\citep{Bergeal:2012fk}, or the scattering parameters of a device under test compared to a reference transmission line~\citep{Yeh:2013bz, Ranzani:2013kl, Ranzani:2013bf}. These methods may require separate cool-downs or multiple switched cryogenic standards, increasing measurement time and uncertainty due to unavoidable change of parameters when the microwave lines are reassembled.
In experiments with superconducting qubits or resonators, some physical effect specific to the  circuit is often used for calibration purposes.
For example, photon numbers have been accurately calibrated through the cross-Kerr effect~\citep{Hoi:2013dh} or via the Stark shift of a qubit-cavity system~\citep{Schuster:2005fv, Schuster:2007ki}. The latter has been extended to multi-level quantum systems (qudits) to deduce the unknown signal frequency and amplitude from the higher level AC Stark shift~\citep{Schneider:2018tw}.
Another method uses a phase qubit as a sampling oscilloscope by measuring how the flux bias evolves in time~\citep{Hofheinz:2009tt}. Other approaches are suitable for correcting pulse imperfections~\citep{Gustavsson:2013go, Bylander:2009ke}. 
An interesting recent proposal uses a transmon qubit coupled to a readout resonator to characterise qubit control lines in the range of 8 to 400 MHz in situ. Unfortunately it is limited by the decoherence time of the qubit~\citep{Jerger:2017wj}.

In this letter, we present a quantum sensor of absolute power operating in the microwave range and at cryogenic temperatures based on a two-level system in a transmission line.
This sensor measures the photon rate (propagated radiation) in a wide frequency range by tuning the two-level system. Importantly, the sensor itself does not disturb the transmission line when detuned.
The sensor can be inserted as an additional lossless element into the transmission line close to the reference plane of another device of interest or used for calibration of transmission lines, microwave components or devices within dilution refrigerators. The working principle is independent of the two-level system used, its implementation and dephasing to first order.
We implement the absolute power quantum sensor using a superconducting flux qubit~\citep{Mooij} strongly coupled to a one-dimensional transmission line~\citep{Astafiev:2007gd, Astafiev:2010cc, Astafiev:2010cm, Hoi:2013dh, Hoi:2013dr}, but in principle it can be implemented with any two-level system as long as it satisfies the strong coupling condition.
We demonstrate several methods for measuring the absolute power, the fastest relying on the concept of continuous wave mixing \citep{PhysRevA.100.013808, PhysRevA.98.041801}.
The accuracy of each method is evaluated by comparing the absolute power sensed at the same frequency by four different qubits on the same chip.

\section{Device and working principle}
\begin{figure*}[!htb]
\centering
\includegraphics[width=2\columnwidth]{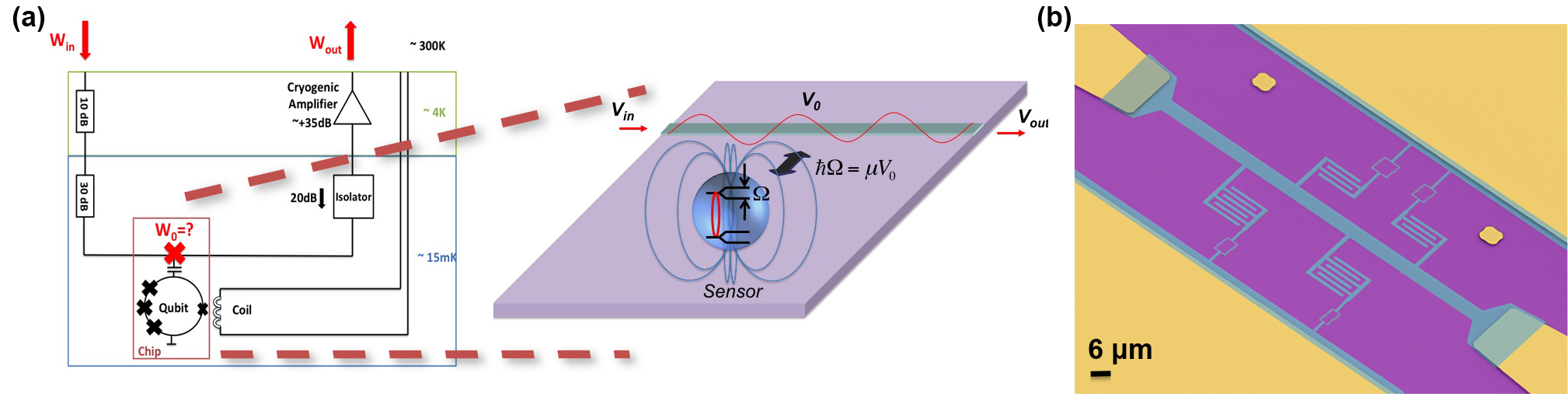}
\caption{a) Schematic of a cryogenic environment together with an illustration of the chip containing a two-level system - the absolute power quantum sensor - coupled to a transmission line. Knowledge of absolute power $W_0$ supplied to a chip at cryogenic temperatures,  are important for most quantum technologies with superconducting circuits. The two-level system with dipole moment $\mu$ interacts with the field $V_{0}$ containing many photons giving rise to coherent oscillations at the Rabi frequency $\Omega$. b) False-coloured SEM image of the sample chip featuring four Al (blue) flux qubits on the undoped silicon oxide substrate (violet) and Au ground planes, markers and bonding pads (yellow).}
\label{fig:1}
\end{figure*}

Our quantum sensor relies on the principle that when a two-level system is illuminated by coherent electromagnetic waves $V_0 e^{-i\omega t}$ with incident photon rate, $\nu$, only a fraction of the incident photons is absorbed with rate $\Omega$. As illustrated in Fig.~\ref{fig:1}(a), the incident  electromagnetic wave couples to the two-level system via the dipole interaction energy, $\hbar\Omega=\mu V_{0}$, where $\mu$ is the dipole moment, and $V_{0}$ is the voltage amplitude of the microwave signal we aim to sense.
The incident photon rate is $\nu=V_{0}^2/(2 Z \hbar \omega)$, where $Z$ is the impedance of the transmission line that guides the microwave photons to the two-level system at angular frequency $\omega$. 

We start with the ideal case of strong coupling of a two-level artificial atom to a 1-D transmission line, where non-radiative relaxation is negligible.
Inserting the expression for the relaxation rate $\Gamma_{1}=(\mu^2\omega Z)/\hbar$~\cite{Astafiev:2010cm, Peng:2016kg} gives
\begin{equation}
\nu=\frac{\Omega^2}{2\Gamma_{1}}.
\label{eq:photonrate}
\end{equation}
To sense the incident power $W_{0}=\nu \hbar \omega$ we need to find two parameters: the Rabi frequency, $\Omega$, and the relaxation rate, $\Gamma_{1}$, (or $\mu$). These two quantities may be measured independently (eg. two separate measurements) as the relaxation rate $\Gamma_1$ (or the dipole moment $\mu$) is a property of the presented sensor whereas the Rabi frequency $\Omega$ relates to the quantity sensed.

We study different methods of finding the required quantities $\Omega$ and $\Gamma_{1}$: (i) by probing the two-level system for reflection through the transmission line, (ii) quantum oscillations, (iii) the Mollow triplet and (iv) wave mixing~\cite{Dmitriev:2017vda}. Note that, Fig.~\ref{fig:1}(a) shows the cryogenic environment only, but each method requires somewhat different experimental set-ups at room temperature. Even though we put effort into keeping the total attenuation similar, there are some variations across the methods.

For this reason, we benchmark our absolute power sensor at $7.48$ GHz to which we can tune each of the four flux qubits with different parameters available for comparison in our device. As seen in Fig.~\ref{fig:1}(b), each flux qubit consists of an Al superconducting loop and four $Al/AlO_{x}/Al$ Josephson junctions fabricated on a silicon oxide substrate, where one of the Josephson junctions, the $\alpha$-junction, has a reduced geometrical overlap by a factor of $\alpha$~\citep{PhysRevA.100.013808}. The coupling capacitance $C_c$ to the 1D transmission line  and $\alpha$-junction was varied; two qubits have been designed to have a coupling capacitance of $C_c=3$ fF with $\alpha=0.5$, while the remaining two qubits have $C_c=5$ fF, $\alpha=0.45$. All qubits have been co-fabricated on one sample chip using electron-beam lithography and shadow evaporation technique with controllable oxidation.

Importantly, with the coupling capacitances $C_c \leq 5$ fF at $7.5$ GHz, the reflection is negligible when the qubits are detuned from the resonance. The reflected power $W_{r}$ on the two-level system from the propagated microwave of power $W_0$ is given by $W_{r}/W_0 = (\omega C_c Z_0/2)^2$, here resulting in $W_{r}  < 3\times 10^{-5} W_0$ and the transmission line can then be regarded as being a low loss and a well matched $50$ $\Omega$ line.

\begin{figure*}[!htb]
\centering
\includegraphics[width=2\columnwidth]{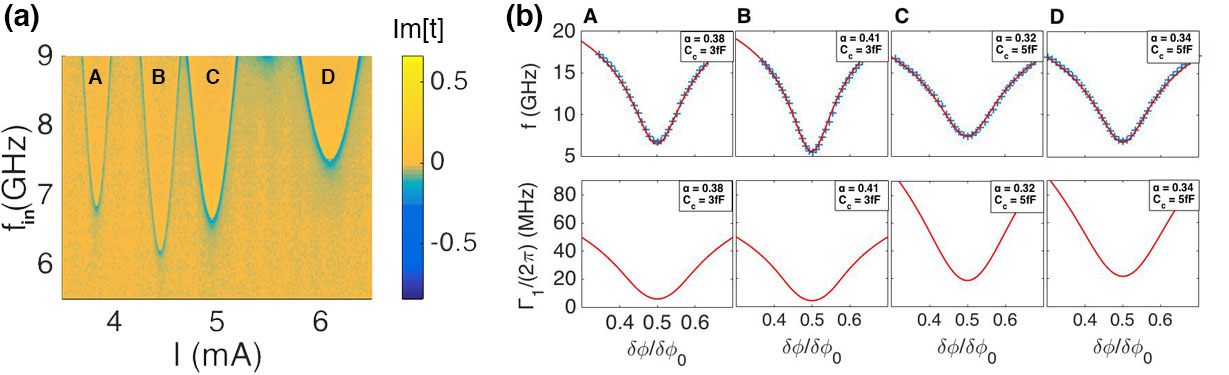}
\caption{a) Transmission spectroscopy, Im$[t]$, of four flux qubits as a function of incident microwave frequency $f_{in}$ and current $I$ through an external superconducting coil providing the bias flux $\phi_b$. The transition frequencies are revealed as dips in the transmission spectrum. We benchmark the absolute power sensor at $7.48$ GHz. b) Transition frequencies $f$ (top row), and relaxation rates (bottom row) $\Gamma_1/(2\pi)$ as a function of flux $\delta\phi/\phi_0$ of artificial atoms A,B,C and D where $\phi_0$ is the flux quantum and $\delta\phi=\phi_b-\phi_0/2$. Blue markers are experimental points taken from transmission spectroscopy. Solid lines are numerical simulations calculated with charging energy $E_{C}/h=(2e)^2/(2 C_J h)=10$ GHz  (with  junction capacitance $C_J$) and Josephson energy $E_J/h = 53$ GHz with $\alpha$ and coupling capacitance $C_c$ shown in the insets.}
\label{fig:Spec}
\end{figure*}

The qubits held at 12 mK are revealed through transmission spectroscopy as seen in Fig.~\ref{fig:Spec}(a). Although, by design, two in four qubits should be identical (apart from their transmission spectrum in magnetic field, since their loop area was varied), a clear spread of energies is visible due to technological limitations. We fit numerical simulations for each qubit to the shape of the transition frequency (Fig.~\ref{fig:Spec}(b)).

To characterise the sensors' relaxation rates we adjust the external field to tune each qubit to $7.48$ GHz.
We drive and readout the qubit with energy splitting $\hbar\omega_a$ using a vector network analyser (VNA). The qubit driven by a microwave tone $V_0 e^{-i\omega t}$ can be described in the rotating wave approximation by the  Hamiltonian $H=\frac{\hbar \delta\omega}{2}\sigma_z - \frac{\hbar\Omega}{2}\sigma_x$, where $\delta\omega=\omega-\omega_a$ is detuning from the resonance of the qubit and $\sigma_{x,y,z}$ are the Pauli matrices. The dynamics of the system is well described by the master equation $\dot{\rho}=-\frac{i}{\hbar}[H,\rho]+\hat{L}[\rho]$ with the Lindblad term $\hat{L}[\rho]=-\Gamma_{1}\sigma_{z}\rho_{11}-\Gamma_{2}(\sigma^+ \rho_{10} +\sigma^- \rho_{01})$ where $\Gamma_{2}$ is the dephasing rate.
When the artificial two-level atom is driven close to its resonance, it acts as a scatterer and generates two coherent waves propagating forward and backward with respect to the driving field~\citep{Astafiev:2010cm}
\begin{equation}
V_{sc}(x,t)=i\frac{\hbar\Gamma_{1}}{\mu}\braket{\sigma^-}e^{ik|x|-i\omega t},
\label{eq:Vsc}
\end{equation}
where $\braket{\sigma^-}=\rho_{10}$ is found from the stationary solution of the master equation.
We measure transmission coefficients $t=1+V_{sc}/V_0$, where $V_0$ and $V_{sc}$ are voltage amplitudes of the incident and scattered electromagnetic waves respectively ~\citep{Astafiev:2010cm, Hoi:2013dr}. We detect the qubit resonances as a sharp dip in the power transmission coefficient $|t|^2$, and reach a power extinction $(1-|t|^2)>85\%$ for all qubits at $7.48$ GHz, confirming strong coupling to the transmission line. In what follows we further assume that the relaxation rate $\Gamma_1$ is dominated by the radiative relaxation to the transmission line, an assumption that is justified in the strong coupling regime.

The reflection coefficient is defined as $V_{sc}=-rV_{0}$, using the relation $r=1-t$ and Eq.~\ref{eq:Vsc} we have~\citep{Astafiev:2010cm}
\begin{equation}
r=\frac{\Gamma_{1}}{2\Gamma_{2}}\frac{1+i\delta\omega/\Gamma_{2}}{1+(\delta\omega/\Gamma_{2})^2+\Omega^2/\Gamma_{1}\Gamma_{2}}.
\label{eq:r}
\end{equation}
As seen in Eq.~\ref{eq:r} (and Fig.~\ref{fig:Br}(a)) the peak in reflection becomes insensitive to driving power at weak driving powers. Fitting the reflection curve  in this limit of low driving power, we find the dephasing $\Gamma_2$ and radiative relaxation rate $\Gamma_1$, which are in good agreement with the numerical simulations of each qubit. Results are tabulated in Table.~\ref{table:Calibrator} where the quoted uncertainties (one standard deviation) of $\Gamma_1$ and $\Gamma_2$ are deduced from the covariance matrix of the fit to the data. Other sources of errors include normalisation errors or drifts in frequency due to qubit instability.
Frequency fluctuations in state-of-the-art-qubits are typically on the scale of kHz~\citep{Burnett:2019kh, Schlor:2019vc}. Here, the qubit linewidths are several MHz and the contribution of frequency fluctuations of the qubit to the lineshape is thus expected to be negligible. 
Likewise frequency variations due to instabilities in the flux bias also remain negligible as we do not observe any increase in fluctuations for the qubits operated away from their degeneracy points.
We normalise the transmission around the qubit resonance by the transmission away from the qubit resonance. This requires tuning the external magnetic field. Another likely source of error are temporal variations in the power generated and measured by the VNA. This was independently measured to vary $\pm0.25$ dB over an hour (the typical timescale for measurements). This translates to a relative uncertainty in $\Delta r/r = 0.02$ and $\Delta\Gamma_1/\Gamma_1$ of $0.03$.

\begin{table}[!h]
\centering
\begin{tabular}{|c|c|c|c|c|c|}\hline
Qubit & $\omega_0/2\pi$ [GHz]   & $1-|t|^2$ & $\Gamma_1/2\pi$ [MHz] & $\Gamma_2/2\pi$ [MHz]\\ \hline
A & $6.83$  & $92\%$ & $8.2 \pm 0.2$ & $5.7 \pm 0.1$ \\
B  & $6.19$ & $87\%$ & $7.8 \pm 0.2$ & $6.2 \pm 0.1$ \\
C & $6.63$ & $93\%$ & $16.4 \pm 0.4$ & $10.4 \pm 0.2$ \\
D & $7.46$ & $94\%$ & $18.4 \pm 0.3$ & $11.6 \pm 0.1$ \\
 \hline
\end{tabular}
\caption{Transition frequency $\omega_0/2\pi$ (at $\delta\phi/\phi_0=0.5$), power extinction $1-|t|^2$, relaxation and dephasing rates $\Gamma_1/2\pi$ and $\Gamma_2/2\pi$ respectively at $\omega_a/2\pi=7.46$ GHz (ie. $\delta\phi/\phi_0\neq 0.5$ for qubits A, B, and C)  of the four flux qubits (denoted as A, B, C, and D).}
\label{table:Calibrator}
\end{table}

\section{Methods}
Having characterised the four sensors, we now present different methods of measuring the quantity that relates to the power. Due to slightly different experimental setups required at room temperature we arrive at somewhat different attenuation values observed across the different methods. We verify that the measured attenuation and gain is in reasonable agreement with the transmission measured through the cryostat at room temperature.

\subsection{Reflection in the transmission line}
\label{reflection}
At $\delta\omega=0$ the reflection coefficient simplifies to $r=\Gamma_1^2/(2\Gamma_1\Gamma_2 + 2\Omega^2)$, and substituting the photon rate (Eq.~\ref{eq:photonrate}) gives
\begin{equation}
W_{0}=\bigg(\frac{\Gamma_1}{4 r}-\frac{\Gamma_2}{2}\bigg)\hbar\omega.
\label{eq:r_abspower}
\end{equation}

\begin{figure}[!htb]
\centering
\includegraphics[width=1.1\columnwidth]{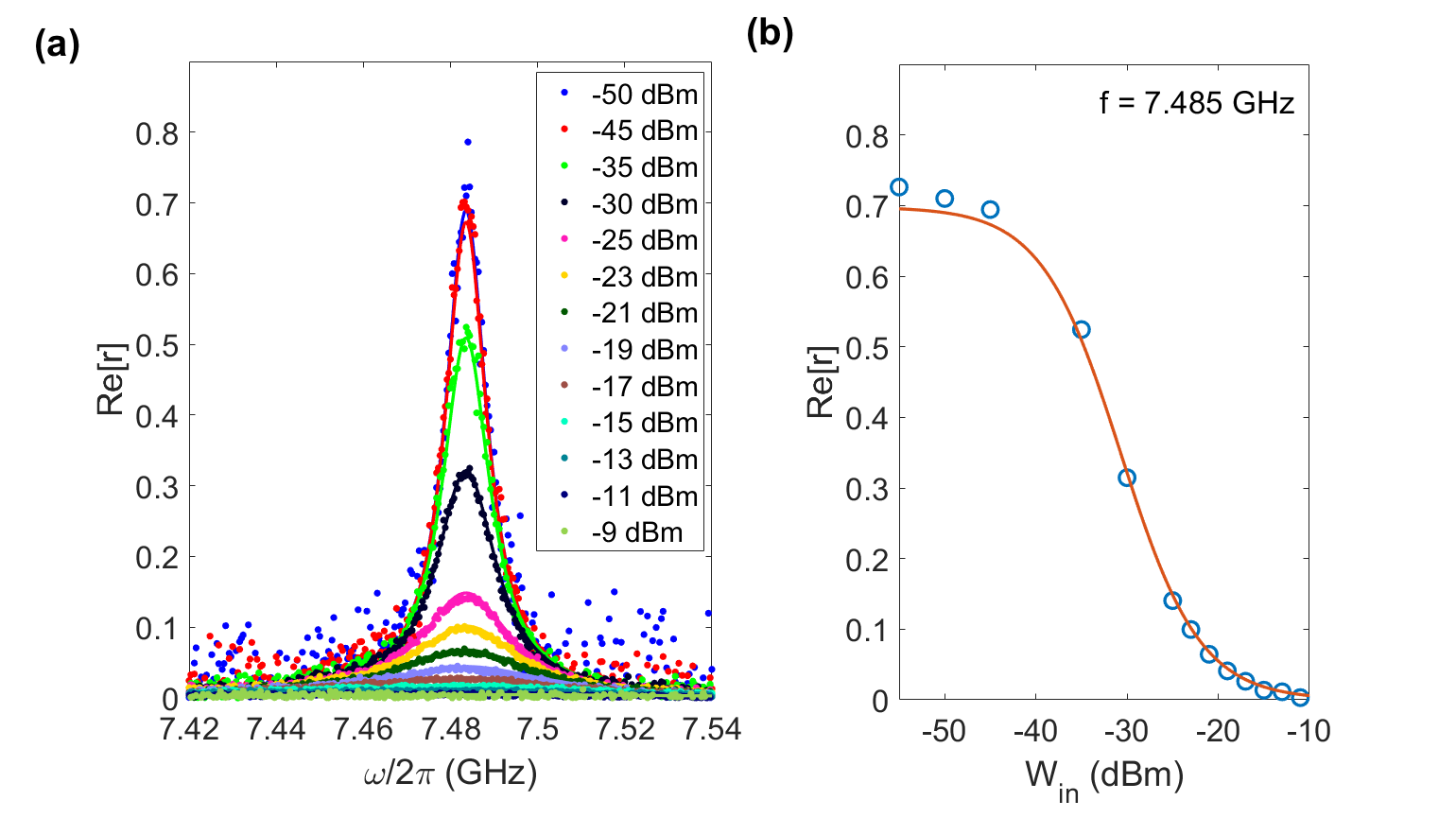}
\caption{Qubit A. a) Reflection as function of frequency for a set of input powers $W_{in}$. Markers are experimental data, solid lines are fits to Eq.~\ref{eq:r}. b) Reflection at $f=7.485$ GHz versus input powers $W_{in}$. Markers are experimental data and solid line is a fit.}
\label{fig:Br}
\end{figure}

Using a VNA, we measure transmission $t$ around $7.48$ GHz for a range of generator input powers $W_{in}$ and deduce the reflection via $r=1-t$ for all four qubits as a function of frequency (Fig.~\ref{fig:Br}(a)).

In Fig.~\ref{fig:Br}(b)), we plot $\rm{Re}(r)$ at $\delta\omega=0$ versus generator input powers $W_{in}$ and fit this curve to Eq.~\ref{eq:r} with $\Omega^2=k W_{in}$ as the only fitting parameter, where $k$ is an attenuation constant relating the input power to the Rabi frequency.
We then calculate the absolute power $W_0$ according to Eq.~\ref{eq:photonrate} (mutliplied by $\hbar\omega$). 
To propagate errors, we take the uncertainty of $\Gamma_1$ and $\Omega$ from the fits (see Table~\ref{table:Calibrator} and Fig.~\ref{fig:Br}(b) respectively). These provide the main source of error in sensing the power. Fluctuations in qubit parameters give a negligible contribution.

\begin{figure}[!htb]
\centering
\includegraphics[width=1\columnwidth]{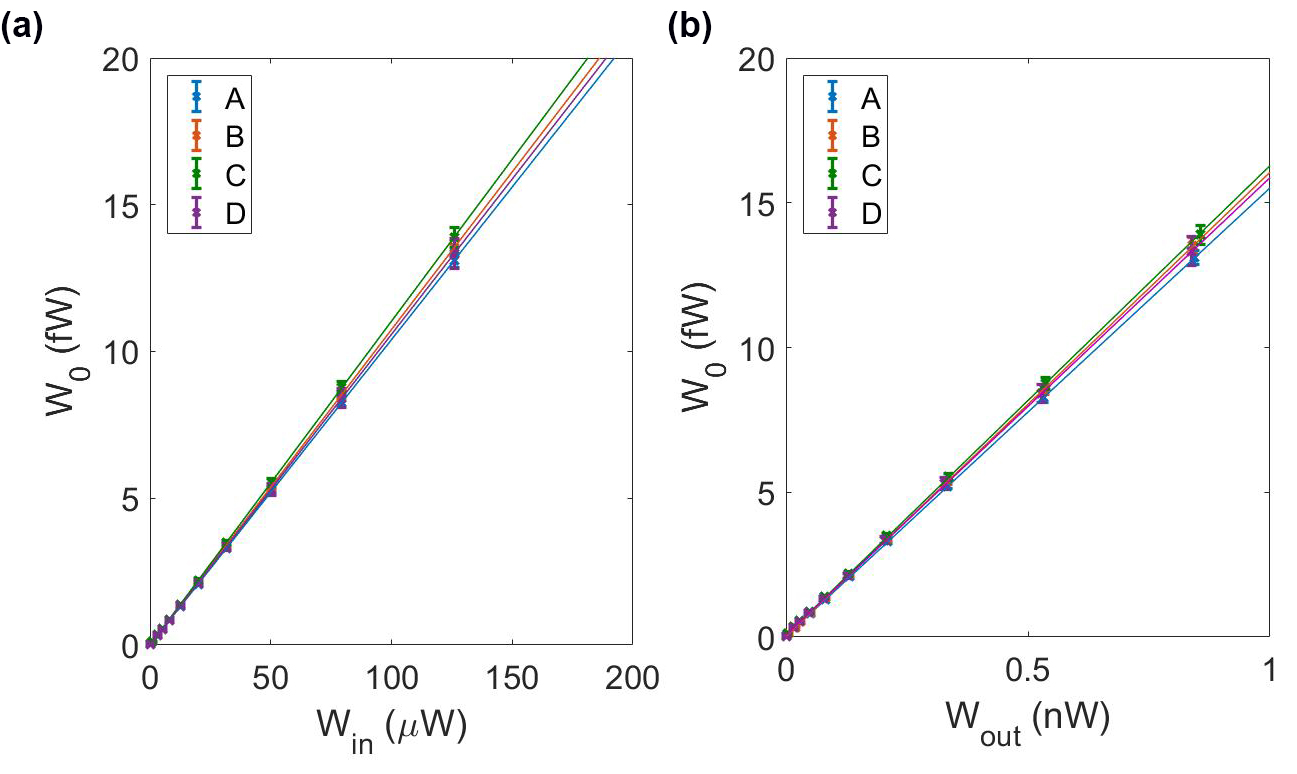}
\caption{The absolute power $W_0$ sensed by qubits A, B, C, and D (Table~\ref{table:Calibrator}) at 7.48 GHz as a function of  (a) input power $W_{in}$ and as a function of (b) output power $W_{out}$. The slope of the linear fits (solid lines) represent (a) attenuation and (b) gain in our measurement circuit.}
\label{fig:r-sum}
\end{figure}

In Fig.~\ref{fig:r-sum} we plot the absolute power $W_0$ sensed by qubits A, B, C, and D against $W_{in}$ ($W_{out})$ where the slope represents the attenuation (gain) in our system. We fit this slope for each qubit (see solid lines in Fig.~\ref{fig:r-sum}) and find that the obtained attenuation and gain coefficients, listed in Table~\ref{table:smith}, are in agreement within $\pm0.2$~dB, which is comparable to the temporal variations in the measured $S_{21}$ of the VNA, indicating that there is no significant device dependent systematic error present. Further, the result is consistent with the expected attenuation of approximately $100$ dB in this particular measurement set-up: In the input line we had placed $90$ dB of attenuators and the coaxial wiring is expected to add roughly $10$ dB in attenuation, as verified at room temperature.
\begin{table}[h!]
\centering
\begin{tabular}{|c|c|c|}\hline
Qubit & Attenuation [dB] & Gain [dB] \\ \hline
A & $-99.8\pm0.2$  & $48.1\pm0.2$\\ 
B & $-99.7\pm0.3$ & $48.0\pm0.3$ \\
C & $-99.6\pm0.4$ & $47.9\pm0.4$ \\
D & $-99.8\pm0.4$ & $48.0\pm0.4$\\ \hline
combined & $-99.8\pm0.2$ &$48.0\pm0.2$\\ \hline
\end{tabular}
\caption{Gain and attenuation coefficients obtained from qubits A, B, C, and D for the reflection through the transmission line method. Errors were propagated from the uncertainties in $\Gamma_1$ as listed in Table~\ref{table:Calibrator} and the uncertainties in $\Omega$ as extracted from the fit of $r$ as a function of input power $W_{in}$.}
\label{table:smith}
\end{table}

It shall be noted that some power may leak from input to output of the chip via ground planes or box modes. This power can interfere with the signal resulting in distortions in the measured reflection curve, or become apparent as an offset which is subtracted when fitting experimental points in Fig.~\ref{fig:Br}(b).

\subsection{Rabi oscillations}
\label{Rabi}
\begin{figure}[!h]
\centering
\includegraphics[width=1\columnwidth]{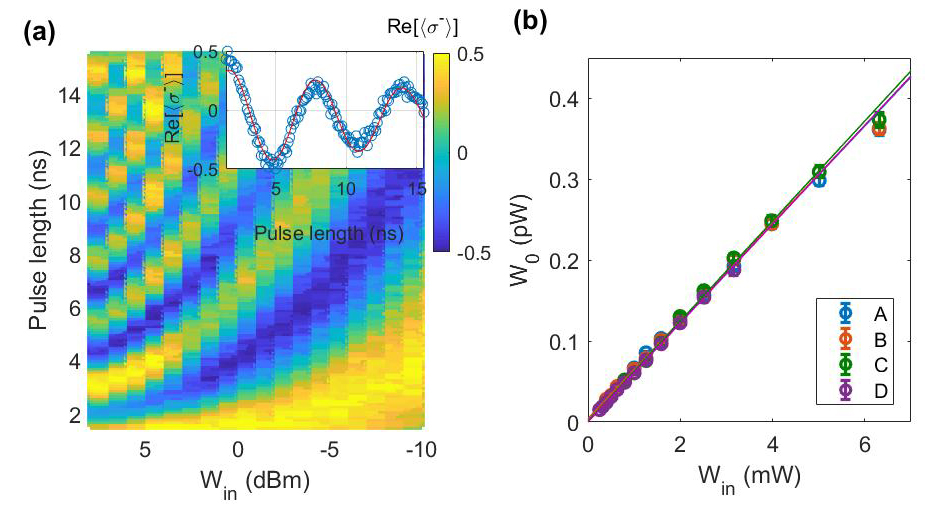}
\caption{(a) Rabi oscillations (of qubit B) for input powers $W_{in}$ ranging from -10 to 8 dBm. The inset shows the normalised dipole moment $\langle\sigma^-\rangle$ for an input power of $W_{in}=-2$ dBm. (Blue circles are experimental data, red solid line is a linear fit.) (b) The absolute power $W_0$ sensed by qubits A, B, C, and D (Table~\ref{table:Calibrator}) at 7.48 GHz as a function of input power $W_{in}$. The slope of the linear fits (solid lines) represent attenuation in our measurement circuit.}
\label{fig:BQosc}
\end{figure}

An alternative method comprises measuring $\Omega$ directly and deducing the absolute power via $W_0=\hbar\omega\Omega^2/(2\Gamma_1)$. We obtain $\Omega$ for a set of driving powers $W_{in}$ by modifying the measurement circuit and performing quantum oscillation measurements. At the input, an incident microwave pulse is formed with varying pulse length from 1.5 ns to 15.5 ns to excite the qubit.
We perform Rabi oscillation measurements for all qubits tuned to 7.48 GHz for a range of input microwave powers, $W_{in}$, set at the microwave generator at room temperature. For each input power, we extract the period from fits to the measured Rabi oscillations (Fig.~\ref{fig:BQosc}(a)). As expected, we observe a linear relationship between Rabi frequency and driving amplitude. From this fit, we find that the typical uncertainty on the deduced Rabi frequency is $\pm10$ MHz. This combined with the uncertainty in $\Gamma_1$ are the main sources of error in the measured absolute power.

Fig.~\ref{fig:BQosc}(b) shows the absolute power $W_0$ sensed by qubits A, B, C, and D (Table~\ref{table:Calibrator}) as a function of input power $W_{in}$. We fit the slope for each qubit individually and find a spread of $0.1$ dB in the obtained attenuation coefficients. We expect the mixers and filters that were added to the experimental set-up for the creation of the excitation pulse to contribute around $2.5$ dB, measured at room temperature. Taking this additional attenuation into account, the obtained attenuation coefficients are also in agreement with the ones extracted using the previous method.
\begin{table}[h!]
\centering
\begin{tabular}{|c|c|}\hline
Qubit & Attenuation [dB] \\ \hline
A & $(-102.2\pm0.4)$  \\ 
B & $(-102.2\pm0.3)$ \\
C & $(-102.1\pm0.4)$  \\
D & $(-102.1\pm0.2)$ \\ \hline
combined & $(-102.1\pm0.2)$ \\ \hline
\end{tabular}
\caption{Attenuation coefficients extracted from measuring Rabi oscillations of qubits A, B, C and D. The uncertainties were obtained by propagating the error in $\Gamma_1$ as listed in Table~\ref{table:Calibrator} and the uncertainties in $\Omega$ as extracted from the fit of the Rabi Oscillations, which constitute the main sources of error for this method.}
\label{table:Rabi}
\end{table}

A disadvantage of this method is that the measurement time of Rabi oscillations is limited by dephasing and that the combinations of mixers forming the pulse can exhibit non-linear behaviour.  At high input powers the oscillations may distort due to interference with leaked power. It may then become necessary to record the power leakage detuned from the qubit to subtract the background, doubling the already long total measurement time. At relatively low input powers it may not be possible to measure many periods, and the Rabi frequency has to be deduced through linear interpolation.

\subsection{Mollow triplet}
\label{Mollow}
\begin{figure}[!htb]
\centering
\includegraphics[width=1.05\columnwidth]{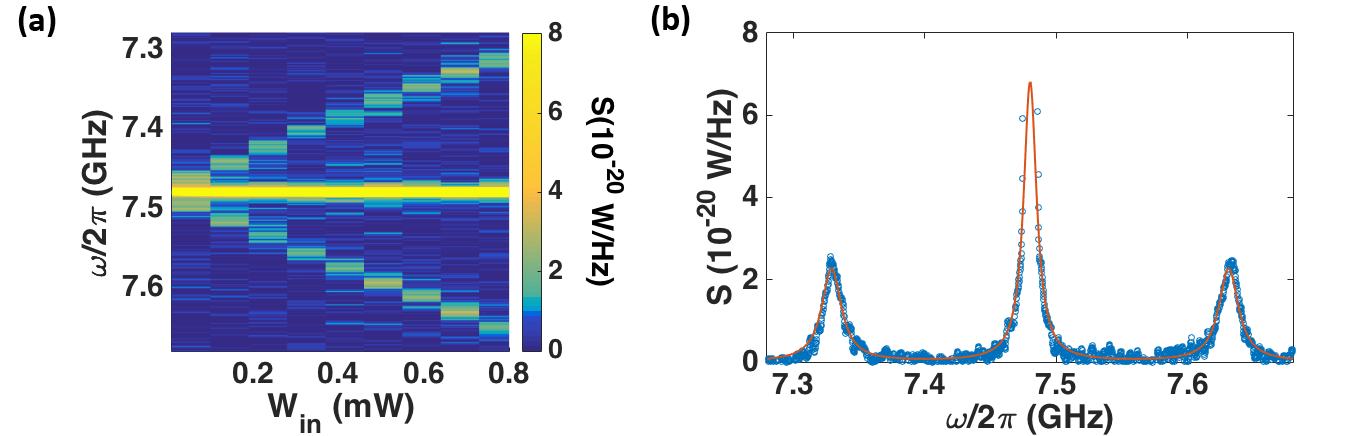}
\caption{a) Mollow triplet (of qubit B) as a function of $W_{in}$ and frequency. b) Linear frequency spectral density of emission power under a resonant drive with fixed driving power $W_{in}=0.73$ mW forming the Mollow Triplet. Experimental data is presented by blue circles. The red solid curve presents the fit of the emission spectrum according to Eq.\ref{eq:mollow} with $\Gamma_1=7.8$ MHz, $\Gamma_2=6.2$ MHz (as in Table~\ref{table:Calibrator}). From the fitting parameters we obtain the Rabi frequency as a function of the external input power.}
\label{fig:BMollow}
\end{figure}

A more robust way to deduce the Rabi frequency $\Omega$ is to measure the artificial atom's incoherent spectrum under strong drive.
The two-level system coupled to a strong driving field ($\Omega^2 \gg \Gamma_{1}^2$) can be described by the dressed-state picture in which the atomic levels are split by $\Omega$. Four transitions between the dressed states are allowed giving rise to the Mollow or resonance fluorescence triplet~\cite{Mollow:1969dd, Baur:2009ee, Astafiev:2010cm}. The side peaks of the triplet are separated by $2\Omega$.
To observe the Mollow triplet we measure the power spectrum around 7.48 GHz using a spectrum analyser under a strong resonant drive (Fig.\ref{fig:BMollow}). To resolve the side peaks we rely on many averages, making this the slowest method. The expected spectral density of the incoherent emission is~\cite{Astafiev:2010cm}
\begin{equation}
\begin{split}
S(\omega)\approx\frac{1}{2\pi}\frac{\hbar\omega\Gamma_{1}}{8}\Big(&\frac{\gamma_{s}}{(\delta\omega+\Omega)^2+\gamma_{s}^2}+\\
&\frac{2\gamma_{c}}{\delta\omega^2+\gamma_{c}^2}+\frac{\gamma_{s}}{(\delta\omega-\Omega)^2+\gamma_{s}^2}\Big),
\end{split}
\label{eq:mollow}
\end{equation}
where half-width of the central and side peaks are $\gamma_{c}=\Gamma_{2}$ and $\gamma_{s}=(\Gamma_{1}+\Gamma_{2})/2$, respectively.

\begin{figure}[!htb]
\centering
\includegraphics[width=1\columnwidth]{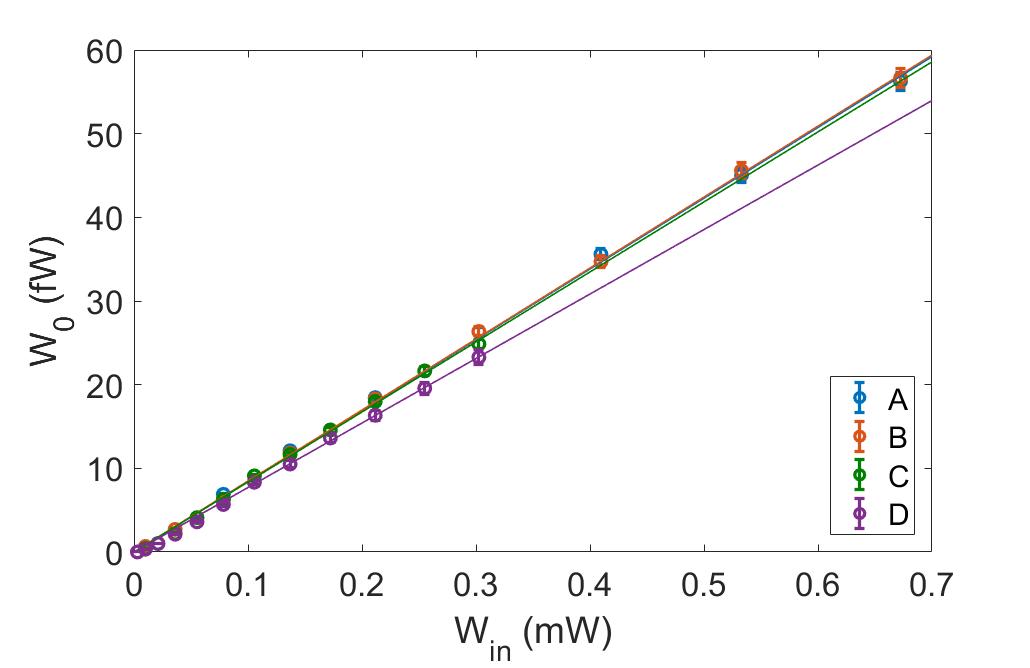}
\caption{The absolute power $W_0$ sensed using method~\ref{Mollow} by qubits A, B, C, and D (Table~\ref{table:Calibrator}) at 7.48 GHz as a function of input power $W_{in}$. The slope of the linear fit (solid lines) represent attenuation in our measurement circuit.}
\label{fig:Mcal}
\end{figure}

We deduce $\Omega$ from fitting the resonance fluorescence emissions spectrum. The fit gives a relative uncertainty $\Delta\Omega/\Omega \leq 0.01$, where $\Delta\Omega$ is the uncertainty of $\Omega$.
We calculate the absolute power according to $W_0=\Omega^2/(2\Gamma_1)\hbar\omega$, where we use the relaxation rates as tabulated in Table~\ref{table:Calibrator}. Again, we plot $W_{in}$ against $W_0$ and fit to a straight line for each qubit individually, as seen in Fig.~\ref{fig:Mcal}. The resulting attenuation coefficients are listed in Table.~\ref{table:mollow}. Here, the measurement set-up at room temperature is similar to the one used for the reflection method. We roughly estimate the gain of the output line in our measurement circuit from the amplitude of the Mollow triplet to be $\sim 45$ dB. The main contributing factor to the error bars in Fig.~\ref{fig:Mcal} is the uncertainty of $\Gamma_1$.

\begin{table}[h!]
\centering
\begin{tabular}{|c|c|c|}\hline
Qubit & Attenuation [dB]  & Gain [dB]\\ \hline
A & $(-100.7\pm0.1)$ & $44\pm1$ \\ 
B & $(-100.7\pm0.1)$ & $45\pm1$\\
C & $(-100.8\pm0.1)$  & $45\pm2$\\
D & $(-101.1\pm0.1)$ & $45\pm2$\\ \hline
combined & $(-101.0\pm0.1)$ & $45\pm2$\\ \hline
\end{tabular}
\caption{Line calibration using the Mollow triplet method. The errors were propagated using the uncertainties in $\Gamma_1$ as listed in Table~\ref{table:Calibrator} and the uncertainties in $\Omega$ as extracted from the fit of the resonance fluorescence triplet.}
\label{table:mollow}
\end{table}

\subsection{Wave mixing}
\label{Wmcal}
\begin{figure}[!htb]
\centering
\includegraphics[width=1.05\columnwidth]{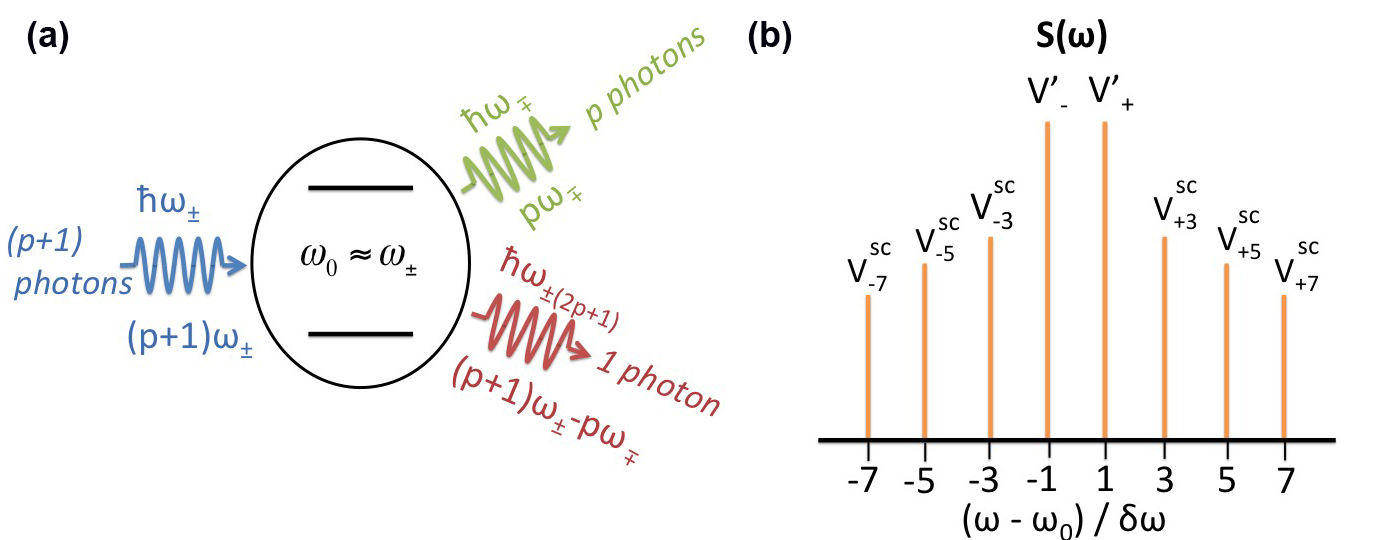}
\caption{a) Schematic of the mixing processes with $2p+1$ interacting photons on a single artificial atom with transition frequency $\omega_0$ resulting in b) spectral components $V_{\pm(2p+1)}^{sc}$ at $\omega_{\pm(2p+1)}=\omega_0 \pm (2p+1)\delta\omega$, where $p\geq 0$ is an integer.}
\label{fig:Schematic-spectrum}
\end{figure}

Some of the methods described above share the potential issue of distortions in the measurements due to interference with leaked power. An elegant solution to this problem is to decouple the input driving powers from the read-out signal in the frequency domain.

We drive the artificial atom by two continuous tones with frequencies $\omega_{-}=\omega_{0}-\delta\omega$ and $\omega_{+}=\omega_{0}+\delta\omega$ where $\omega_{0}=7.48$ GHz and negligible detuning $\delta\omega=5$ kHz $\ll \Gamma_{1}$. The mixing processes can be described in terms of multi-photon elastic scattering. 
For example, a photon at $2\omega_- - \omega_+$ is emitted as a result of absorption of two photons from the $\omega_-$-mode and emission of a single photon from the $\omega_+$-mode. Similarly a photon at $2\omega_+ -\omega_-$ is created due to absorption of two photons from the $\omega_+$-mode and emission of a single photon from the $\omega_-$-mode.
As long as the two driving modes consist of many propagating photons in timescales comparable to relaxation and dephasing rates, $\Gamma_1$ and $\Gamma_2$ respectively, higher-order processes of wave mixing will be present. As illustrated in Fig.~\ref{fig:Schematic-spectrum}, $2p+1$ interacting photons result in spectral components at $\omega_{\pm(2p+1)}=(p+1)\omega_{\pm}-p\omega_{\mp}$, where $p\geq 0$ is an integer. An analytical formula for the amplitude of the scattered spectral components is~\cite{PhysRevA.100.013808}
\begin{equation}
V^{sc}_{\pm(2p+1)}=\frac{(-1)^p \Gamma_1 \tan\theta\tan^p\frac{\theta}{2}}{\Lambda}(V_{\mp}\tan\frac{\theta}{2}-V_{\pm}).
\label{eq:analytical}
\end{equation}
For equal driving amplitudes $\Omega_+=\Omega_-=\Omega$, $\theta=\arcsin\big(\frac{2\Gamma_2\Omega^2}{\Gamma_1|\lambda|^2+2\Gamma_2\Omega^2}\big)$, $\Lambda^{-1}=\frac{\lambda\Gamma_1}{4\Omega^2}$ with $\lambda=\Gamma_2+i\Delta_d$ where $\Delta_d$ is detuning from the central frequency.

We denote the spectral components measured at the frequencies of our driving tones as $V_{\pm}^{'}$ since they consist of the scattered spectral component $V_{\pm}^{sc}$ and the driving amplitude $V_{\pm}$. 

Having already characterised relaxation rates $\Gamma_1$, we only need to record amplitudes of the wave mixing peaks $V_{\pm}^{'}$ and $V^{sc}_{\mp 3}$ for a set of powers $W_{in}$. Since drive and read-out signals are decoupled in frequency in this method, we do not need to measure the power leakage detuned from the qubit to subtract the background, significantly decreasing the total measurement time.

\begin{figure*}[!htb]
\centering
\includegraphics[width=2\columnwidth]{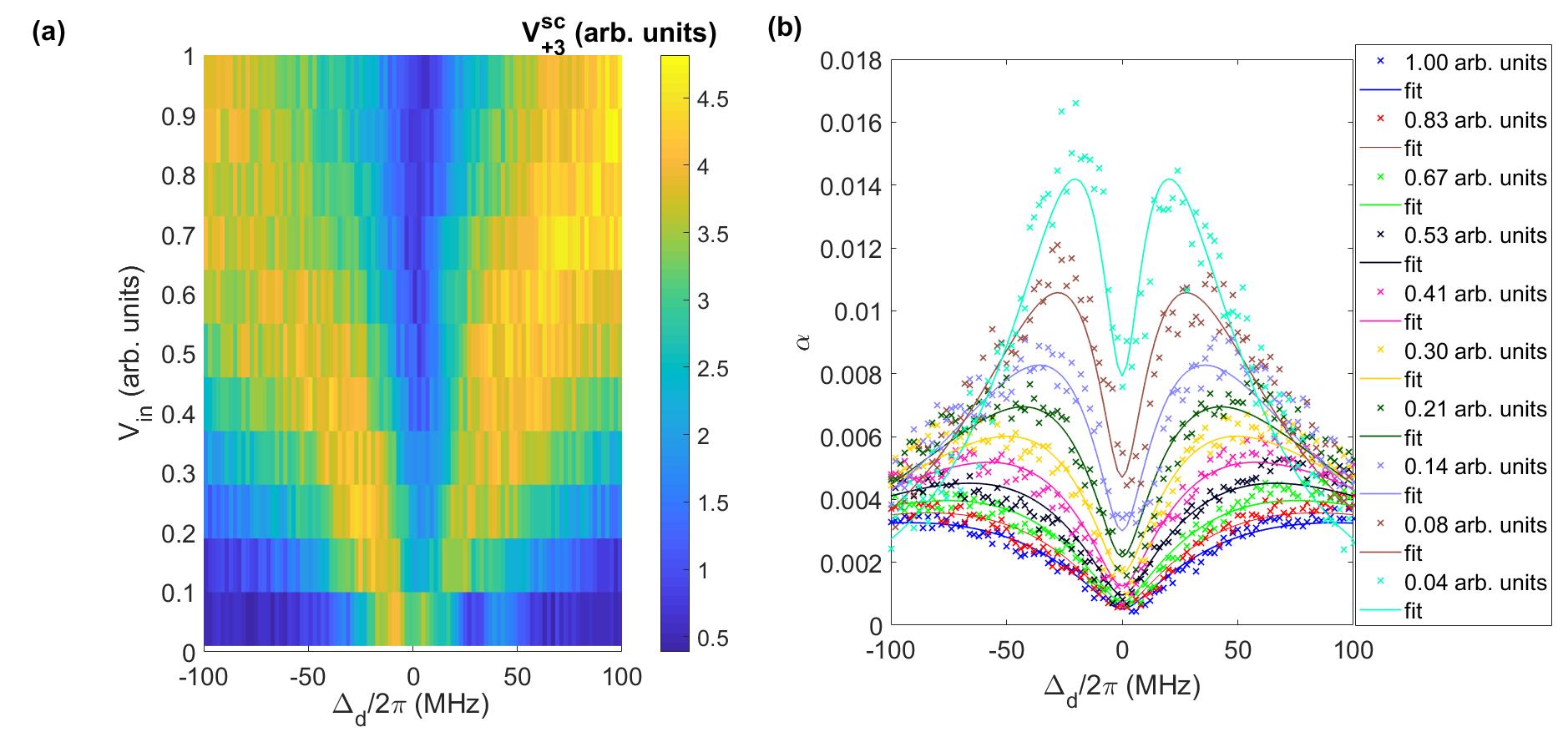}
\caption{Qubit A. a) Autler-Townes-like splitting of the spectral component of the first order ($p=1$) side peak $V^{sc}_{+3}$, that appears at $\omega_{+3}=\omega_0'+3\delta\omega$ due to continuous wave mixing with two drives of equal amplitudes ranging from 1 to 0.1 $(mW)^{1/2}$. b) The ratio $\alpha=V^{sc}_{+3}/V^{'}_{-}$ as a function of detuning $\Delta_d$. The Autlers-Townes-like splitting grows with driving power. Markers are experimental points; solid lines are fits.}
\label{fig:3pBuneq}
\end{figure*}

As seen in Fig.~\ref{fig:3pBuneq}, we measure the spectral components as a function of detuning of the central frequency $\omega_0'=\omega_0\pm\Delta_d$ while keeping $\delta\omega$, the separation between the two drives $\omega_\pm=\omega_0'\pm\delta\omega$, constant and observe an Autlers-Townes-like splitting.
Fitting this splitting to Eq.~\ref{eq:analytical} we extract $\Omega$ and its relative uncertainty $\Delta\Omega/\Omega<0.03$.
We calculate the absolute power $W_0$ according to Eq.~\ref{eq:photonrate}. Again, we propagate the errors using the uncertainties in the relaxation rate $\Gamma_1$, as listed in Table ~\ref{table:Calibrator} and the relative uncertainty $\Delta\Omega/\Omega$ as found from the covariance matrix of the fit to the Autlers-Townes like splitting.

To measure the mixing it is necessary to modify the experimental set-up at room temperature and include a microwave combiner, leading to a slightly higher attenuation of roughly $3$ dB compared to the reflection or Mollow triplet method.
We obtain attenuation and gain coefficients for each sensor with a spread of $0.2$ dB by fitting to a straight line.
\begin{figure}[!htb]
\centering
\includegraphics[width=1\columnwidth]{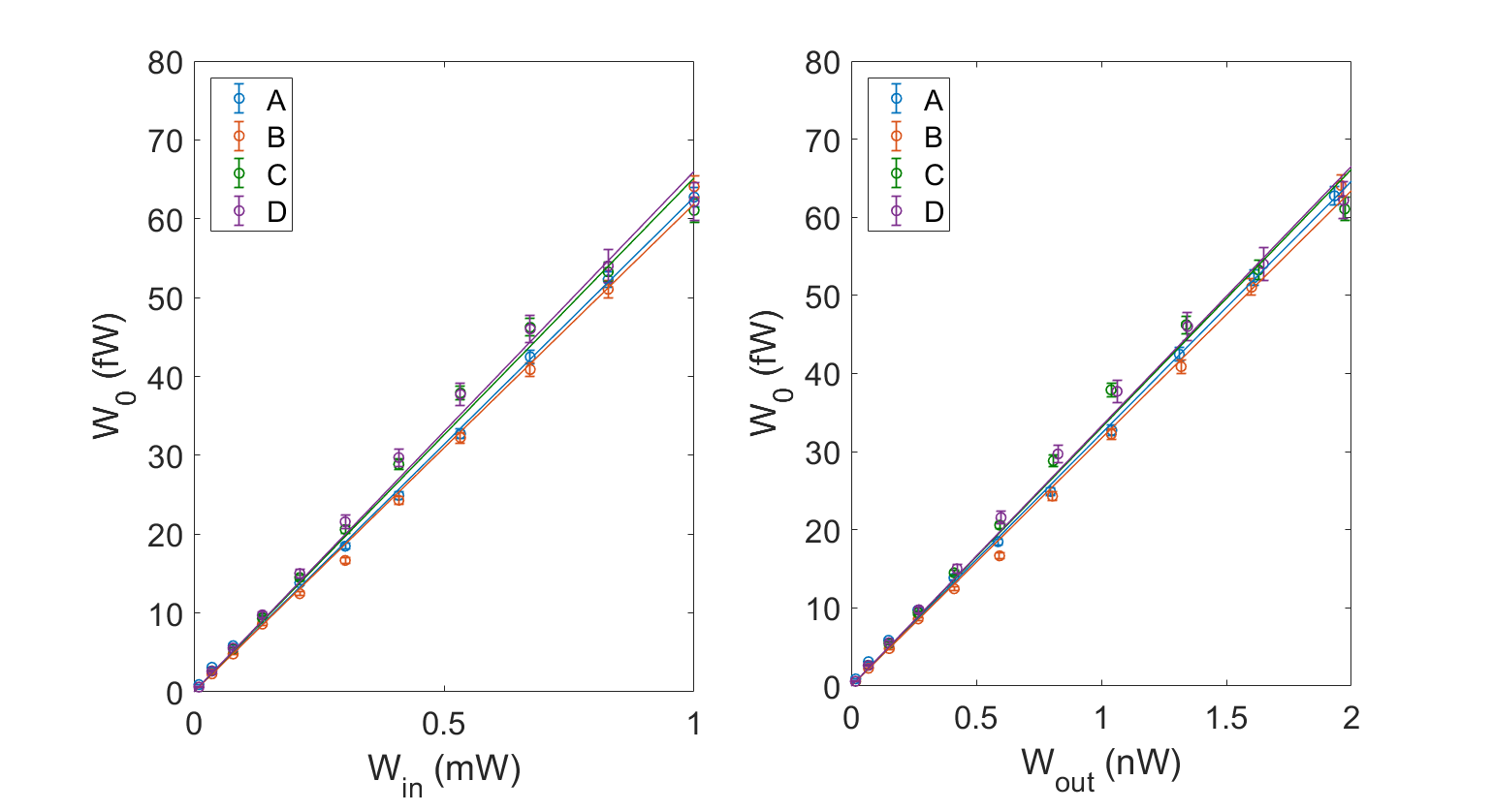}
\caption{Mixing method with equal driving powers: The absolute power $W_0$ sensed by qubits A, B, C, and D (Table~\ref{table:Calibrator}) at 7.48 GHz as a function of  (a) input power $W_{in}$ and as a function of (b) output power $W_{out}$. The slope of the linear fit (solid red line) represents (a) attenuation and (b) gain in our measurement circuit.}
\label{fig:eq-g-a}
\end{figure}

\begin{table}[h!]
\centering
\begin{tabular}{|c|c|c|}\hline
Qubit & Attenuation [dB] & Gain [dB]\\ \hline
A & $-102.0\pm0.1$ & $44.9\pm0.1$ \\ 
B & $-102.1\pm0.2$ & $45.0\pm0.2$ \\
C & $-101.9\pm0.4$ & $44.8\pm0.4$  \\
D & $-101.8\pm0.3$ & $44.8\pm0.3$ \\ \hline
combined & $-102.0\pm0.1$ & $44.9\pm0.1$ \\ \hline
\end{tabular}
\caption{Attenuation coefficients extracted from measuring wave mixing of qubits A, B, C and D. The errors were propagated using the uncertainties in $\Gamma_1$ as listed in Table~\ref{table:Calibrator} and the uncertainties in $\Omega$ as extracted from the fit of the Autler-Townes-like splitting of the spectral components of the first order side peak (Fig.~\ref{fig:eq-g-a}(b)).}
\label{table:mixing}
\end{table}
Finally, the total measurement time can be significantly decreased by measuring a single slice at $\Delta_d=0$.
We introduce the following variables: $\alpha_m=V^{sc}_{\mp3}/V^{'}_{\pm}$ and $x=\Gamma_1/\Omega$ at the exact resonance when $\Delta_d=0$. We set $\Gamma_2/\Gamma_1=\chi$, which reaches minimal value 1/2 in the absence of pure dephasing. Using the variables, we now express the photon emission rate as $\nu=\frac{\Gamma_1}{2}x^{-2}$.
With Eq.~\ref{eq:analytical} we expand $\alpha_m$ in series: $\alpha_m=\frac{x^2}{4}+o(x^3)$, and therefore $x^{-2}=\frac{1}{4\alpha_m}+o(\alpha_m^{\frac{3}{2}})$ and $\nu=\frac{\Gamma_1}{8\alpha_m}\eta$, where $\eta=1-3\sqrt{\chi\alpha_m}+(1-\frac{\chi}{2})\alpha_m+o(\alpha_m^{3/2})$. To first order the photon rate, 
\begin{equation}
\nu_1=\frac{\Gamma_1}{8}\frac{V^{'}_{\pm}}{V^{sc}_{\mp3}},
\label{eq:v1}
\end{equation}

does not contain a dephasing term ($\chi$). To correct for higher orders $\nu_1$ is multiplied by the correction term $\eta$, such that $\nu=\nu_1\eta$.

For example, for $\alpha_m = 10^{-4}$, the approximation of Eq.~\ref{eq:v1} gives the result with an accuracy of $\sim\sqrt{\alpha_m}$, which is about 2\%. Accounting the correction terms  will reduce the derivation error down to $\sim\alpha_m^{3/2}\approx10^{-6}$.
Using this simplified method we can arrive at attenuation and gain values with errors of the same magnitude as the other methods within minutes, greatly speeding up the total measurement time.

\section{Conclusion}

To summarise, we have developed an absolute power quantum sensor based on a superconducting qubit operating in a wide gigahertz range at millikelvin temperatures. Our work addresses the current lack of devices optimised for low-temperature microwave calibration. 
\\ We have shown that the absolute power is determined by two quantities only, the Rabi frequency $\Omega$ and device-dependent relaxation rate $\Gamma_1$. The presented methods are based on measuring spectra of scattered radiation through a transmission line, however, the fastest and most promising technique, from our point of view, relies on a recently demonstrated effect of wave mixing on a quantum system.
For each method, we find that the power sensed by different qubits with different relaxation rates are in agreement. We do not see qubits with similar relaxation rate group, ruling out significant systematic errors in the measurement of the relaxation rate.
We analyse our results for the attenuation and gain in our measurement set-up with a spread smaller than $0.4$ dB across all methods and find that they are in good agreement with our expectations. Table~\ref{table:Calibrator-results-summary-scaled} shows a comparison of the average attenuation for each method scaled to the setup of the reflection method which have a spread of less than $1.4$ dB giving an upper limit to any systematic error of any method.

\begin{table}[]
\centering
\begin{tabular}{|c|c|c|c|c|}\hline
Method & Attenuation [dB]   & Gain [dB] \\ \hline
Reflection (sec.~\ref{reflection}) & $(-99.8\pm0.2)$  & $(48.0\pm0.2)$ \\
Rabi osc. (sec.~\ref{Rabi})  & $(-99.6\pm0.5)$ & - \\
Mollow triplet (sec.~\ref{Mollow}) & $(-101.0\pm 0.1)$ & $(45 \pm 2)$ \\
Mixing (sec.~\ref{Wmcal}) & $(-99.0\pm 0.5)$ & $(44.9\pm0.1)$ \\
 \hline
\end{tabular}
\caption{Summary of attenuation and gain coefficients of the input and output microwave lines in our dilution refrigerator obtained by the different methods and scaled to the reflection setup parameters.}
\label{table:Calibrator-results-summary-scaled}
\end{table}

Our sensor does not affect the transmission of microwaves when detuned in frequency, enabling its use in combination with other microwave devices and allowing the sensor to be incorporated on chip or plugged into the transmission line at a point of interest.
We expect this to be useful for applications in quantum information processing, as well as for fundamental research applications in cryogenic environments.

\section{Acknowledgements}
We thank A. Dmitriev, J. Burnett, T. Lindstr\"{o}m, S. Giblin and A. Tzalenchuk for helpful discussions. 
We gratefully acknowledge the UK Department of Business, Energy and Industrial Strategy (BEIS), the Industrial Strategy Challenge Fund Metrology Fellowships and the Russian Science Foundation (grant N 16-12-00070) for supporting the work.

\newpage
\end{document}